\documentclass[prc,twocolumn,showpacs,showkeys,superscriptaddress]{revtex4}
\usepackage{dsfont}
\usepackage{graphicx,amsmath}
\def\be{\begin{eqnarray}}
\def\ee{\end{eqnarray}}
\def\bc{\begin{center}}
\def\ec{\end{center}}

\def\rmF{{\rm F}}
\def\rmd{{\rm d}}
\def\om{\omega}

\def\half{{\textstyle \frac12}}
\def\quart{{\textstyle \frac14}}
\newcommand{\lsim}{\stackrel{\scriptstyle <}{\phantom{}_{\sim}}}
\newcommand{\gsim}{\stackrel{\scriptstyle >}{\phantom{}_{\sim}}}

\DeclareMathOperator{\Tr}{Tr}
\begin{document}
\title{
Pomeranchuk instability and Bose
condensation  of scalar quanta in a  Fermi liquid }
\author{E.\ E.\ Kolomeitsev}
\affiliation{Matej Bel University, SK-97401 Banska Bystrica, Slovakia}
\author{D.\ N.\ Voskresensky}
\affiliation{ National Research Nuclear University (MEPhI), 115409 Moscow, Russia}
\begin{abstract}
We study excitations in a normal Fermi liquid with a local scalar interaction.
Spectrum of bosonic scalar-mode
excitations is investigated for various values and momentum
dependence of the scalar Landau parameter $f_0$ in the
particle-hole channel. For $f_0 >0$ the conditions are found when
the phase velocity on the spectrum of the zero sound acquires a
minimum at a non-zero momentum. For $-1<f_0 <0$ there are only
damped excitations, and for $f_0<-1$ the spectrum becomes unstable
against a growth of scalar-mode excitations (a Pomeranchuk
instability). An effective Lagrangian for the scalar excitation
modes is derived after performing a bosonization procedure. We
demonstrate that the Pomeranchuk instability may be tamed by the
formation of a static Bose condensate of the scalar modes. The
condensation may occur in a homogeneous or inhomogeneous state
relying on the momentum dependence of the scalar Landau parameter.
Then we consider a possibility of the condensation of the
zero-sound-like excitations in a state with a non-zero momentum in
Fermi liquids moving with overcritical velocities, provided an
appropriate momentum dependence of the Landau parameter $f_0
(k)>0$.
\end{abstract}
\date{\today}
 \maketitle
\section{Introduction}

The theory of normal Fermi liquids was built up by
Landau~\cite{FL}, see in textbooks~\cite{LP1981,Nozieres,GP-FL}.
The Fermi liquid approach to the description of nuclear systems
was developed by Migdal~\cite{Mjump,M67}. In the Fermi liquid
theory the low-lying excitations are described by several
phenomenological Landau parameters. Pomeranchuk has shown in
Ref.~\cite{Pomeranchuk} that Fermi liquids are stable only if some
inequalities on the values of the Landau parameters are fulfilled.

In this work we study low-lying scalar excitation modes
(density-density fluctuations) in the cold normal Fermi liquids
for various values and momentum behavior of the scalar Landau
parameter $f_0$ in the particle-hole channel.  We assume that an
interaction in the particle-particle channel is repulsive and the
system is, therefore, stable against pairing in an s-wave state.
An induced p-wave pairing to be possible at very low temperatures
$T<T_{{\rm c},p}$, see  Refs.~\cite{pwave-pairing}, can be
precluded by the assumption that the temperature of the system is
small but above $T_{{\rm c},p}$.

For $f_0 > 0$ the conditions will be found when the phase velocity
of the spectrum possesses a minimum at a non-zero momentum. This
means that the spectrum satisfies the Landau necessary condition
for superfluidity. As a consequence this may lead to a
condensation of zero-sound-like excitations with a non-zero momentum in
moving Fermi liquids with the velocity above the Landau critical
velocity~\cite{Vexp95}. Similar phenomena may occur in moving
He-II, cold atomic gases, and other moving media, like rotating
neutron stars, cf.
Refs.~\cite{Pitaev84,V93,Melnikovsky,BP12,KV2015,Kolomeitsev:2015dua}.
For $-1<f_0 <0$ excitations are damped, for $f_0<-1$ the spectrum
is unstable against the growth of zero-sound-like modes and
hydrodynamic modes. Up to now it was thought that for $f_0<-1$ the
mechanical stability condition is violated that results in
exponential buildup of the density fluctuations. In hydrodynamic
terms the condition $f_0<-1$ implies that the speed of the first
sound becomes imaginary. This would lead to an exponential growth
of the aerosol-like mixture of droplets and bubbles (spinodal
instability).  In a one-component Fermi liquid spinodal
instability results in a mixed liquid-gas like stationary state
determined by the Maxwell construction if the pressure has a van
der Waals form as a function of the volume. In the isospin
symmetric nuclear matter  (if the Coulomb interaction is
neglected) the liquid-gas phase transition might
occur~\cite{RMS,SVB} for the baryon densities $0.3 n_0\lsim n\lsim
0.7 n_0$, where $n_0$ is the nuclear saturation density. In a
many-component system a mechanical instability is accompanied by a
chemical instability, see Ref.~\cite{Margueron:2002wk}.  The
inclusion of the Coulomb interaction, see
Refs.~\cite{Ravenhall:1983uh,Maruyama:2005vb}, leads to a
possibility of the pasta phase in the neutron star crusts for
densities $0.3 n_0\lsim n\lsim 0.7 n_0$.

The key point of this
work is that we suggest an alternative description of unstable zero-sound-like
modes which might be realized at certain conditions. We
demonstrate that for $f_0<-1$ instability may result  in an
accumulation of  a static Bose condensate of the  scalar field.
The condensate amplitude is stabilized by the repulsive
self-interaction.  The condensation may occur in the homogeneous
either in inhomogeneous state depending on the momentum dependence
of the Landau parameter $f_0(k)$. In the presence of the
condensate the Fermi liquid proves to be stable.

The work is organized as follows. In  Sect.~\ref{sec:sounds} we
study spectrum of excitations in a one-component Fermi liquid in
the scalar channel in dependence on $f_0(k)$.  In
Sect.~\ref{sec:boson} we bosonize the local interaction and
suggest an effective Lagrangian for the self-interacting scalar
modes. In Sect.~\ref{sec:pomer} we study Pomeranchuk instability
for $f_0<-1$ and suggest a novel possibility of the occurrence of
the static Bose-condensate which leads to a stabilization of the
system. In Sect.~\ref{sec:moving} we consider condensation of
scalar excitations in moving Fermi liquids with repulsive
interactions. Concluding remarks are formulated in
Sect.~\ref{sec:conclude}.

\section{Excitations in a Fermi liquid}\label{sec:sounds}

\subsection{Landau particle-hole amplitude.}

Consider the simplest case of a one-component Fermi liquid of
non-relativistic fermions.  As discussed in the Introduction, we
assume that the system is stable against pairing. The
particle-hole scattering amplitude on the Fermi surface obeys the
equation~\cite{LP1981,Nozieres,GP-FL,M67}
\begin{align}
&\widehat{T}_{\rm ph}(\vec{n}\,',\vec{n};q) =
\widehat{\Gamma}^{\om}(\vec{n}\,',\vec{n})
 \nonumber\\
&\quad+ \langle \widehat{\Gamma}^{\om}(\vec{n}\,',\vec{n}'')\,
\mathcal{L}_{\rm ph}(\vec{n}\,'';q)\, \widehat{T}_{\rm
ph}(\vec{n}\,'',\vec{n};q) \rangle_{\vec{n}\,''}\,,
\label{Tph-FL}
\end{align}
where $\vec{n}$ and $\vec{n}\,'$ are directions of fermion momenta
before and after scattering and $q=(\om,\vec{k})$ is the momentum
transferred in the particle-hole channel. The brackets stand for
averaging over the momentum direction $\vec{n}$
\begin{align}
\langle\dots\rangle_{\vec{n}} = \intop\frac{\rmd \Omega_{\vec
n}}{4\,\pi} \big(\dots\big)\,, \label{angl-aver}
\end{align}
and the particle-hole propagator is
\begin{align}
\mathcal{L}_{\rm ph}(\vec{n};q)= \intop_{-\infty}^{+\infty}\!\!
\frac{\rmd \epsilon}{2\pi i}\!\! \intop_{0}^{+\infty}\!\!\frac{
\rmd p\,p^2}{\pi^2} \, {G}(p_{\rmF+})\,{G}(p_{\rmF-})\,,
\end{align}
where we denoted $p_{\rmF
\pm}=(\epsilon\pm\om/2,p_\rmF\,\vec{n}\pm\vec{k}/2)$ and $p_{\rm
F}$ stands for the Fermi momentum. The quasiparticle contribution
to the full Green's function is given by
\begin{align}
G(\epsilon,\vec{p})=\frac{a}{\epsilon-\xi_{\vec{p}}+i\, 0\, {\rm
sign} \epsilon}\,,\quad\xi_{\vec{p}}=\frac{p^2-p_\rmF^2}{2\,m_{\rm
F}^*}\,.\label{Gn-QP}
\end{align}
Here $m^*_{\rm F}$ is the effective fermion mass, and the parameter $a$ determines a quasiparticle
weight in the fermion spectral density, $0<a\leq 1$, which is expressed through the retarded
fermion self-energy $\Sigma_{ \rm F}^R(\epsilon,p)$ as $a^{-1}= 1-(\partial \Re\Sigma_{ \rm
F}^R/\partial \epsilon)_0$. The full Green's function contains also a regular background part
$G_{\rm reg}$, which is encoded in the renormalized particle-hole interaction $\hat{\Gamma}^\om$ in Eq.~(\ref{Tph-FL}).

The interaction in the particle-hole channel can be written as
\begin{align}
\hat\Gamma^{\om}(\vec{n}\,',\vec{n}) =
\Gamma^{\om}_0(\vec{n}\,'\vec{n})\,  \sigma'_0 \sigma_0 +
\Gamma^{\om}_1(\vec{n}\,'\vec{n})\,
(\vec{\sigma}\,'\vec{\sigma})\,.
\label{Gom-fullspin}
\end{align}
 The matrices $\sigma_j$ with $j=0,\dots,3$ act on incoming
fermions while the matrices $\sigma'_j$ act on outgoing fermions;
$\sigma_0$ is the unity matrix and other Pauli matrices
$\sigma_{1,2,3}$ are normalized as ${\rm
Tr}\sigma_i\sigma_j=2\delta_{ij}$. We neglect here the spin-orbit
interaction, which is suppressed for small transferred momenta
$q\ll p_{\rm F}$. The scalar and spin amplitudes in
Eq.~(\ref{Gom-fullspin}) can be expressed in terms of
dimensionless  scalar and spin Landau parameters
\begin{align}
\tilde{f}(\vec{n}\,',\vec{n})
&=a^2\,N_0\Gamma_0^\om(\vec{n}\,',\vec{n})\,, \nonumber\\
\tilde{g}(\vec{n}\,',\vec{n}) &=a^2\,N_0
\Gamma_1^\om(\vec{n}\,',\vec{n})\,, \label{L-param}
\end{align}
where the normalization constant is chosen as in applications to
atomic nuclei~\cite{M67,SaperFayans,MSTV90} with the density of
states at the Fermi surface, $N_0=N(n=n_0)$, taken at the  nuclear
saturation density $n_0$ and $N(n)=\frac{m_{\rm F}^* (n)\,p_{\rm
F}(n)}{\pi^2}$. Such a normalization is at variance with that
used, e.g., in Refs.~\cite{FL,LP1981,Nozieres,GP-FL}. Their
parameters are related to ours defined in Eq.~(\ref{L-param}) as
$f=N \,\tilde{f}/N_0$ and $g=N\,\tilde{g}/N_0$.

The Landau parameters can be expanded in terms of the Legendre
polynomials $P_l(\vec{n}\cdot\vec{n}\,')$,
\begin{align}
\tilde{f}(\vec{n}\,',\vec{n})=\sum_l \tilde{f}_l\, P_l
(\vec{n}\cdot\vec{n}\,')\,,
\end{align}
and the similar expression exists for the parameter $\tilde{g}$.
The Landau parameters $\tilde{f}_{0,1}$, $\tilde{g}_{0,1}$ can be
directly related to observables~\cite{GP-FL}. For instance, the
effective quasiparticle mass
is given by~\cite{LP1981,Nozieres}
\begin{align}
\label{mef}
\frac{m^*_{\rm F}}{m_{\rm F}} = 1+a^2\,N\,\overline{\Gamma_0(\cos\theta)\cos\theta}=
1+\frac13 f_1\,,
\end{align}
where the bar denotes the averaging over the azimuthal and polar
angles. The positiveness of the effective mass is assured by
fulfillment of the Pomeranchuk condition $f_1>-3$. Note that the
traditional normalization of the Landau parameters (\ref{L-param})
depends explicitly on the effective mass $m^*$ through the density
of states $N$. Therefore, it is instructive to rewrite
Eq.~(\ref{mef}) using the definition in Eq.~(\ref{L-param})
\begin{align}
\frac{m^*_{\rm F}}{m_{\rm F}}=\frac{1}{1-\frac13
\frac{m_\rmF}{m^*_{\rmF}(n_0)}\tilde{f}_1}\,.
\end{align}
From this relation we obtain the constraint
$\tilde{f}_1<3\,m^*_{\rmF}(n_0)/m_\rmF$ for the effective mass to
remain positive and finite; otherwise the effective mass tends to
infinity in the point where
$\tilde{f}_1=3\,m^*_{\rmF}(n_0)/m_\rmF$. Thus, for the systems
where one expects a strong increase of the effective mass, the
normalization (\ref{L-param}) of the Landau parameters would be
preferable. A large increase of the effective fermion mass may be
a sign of a quantum phase transition dubbed in
Refs.~\cite{KhodelShaginian,Khodel:2011dx} as a fermion
condensation. The latter is connected with the appearance of
multi-connected Fermi surfaces. As demonstrated in
Ref.~\cite{Voskresensky:2000px} if such a phenomenon occurs in
neutron star interiors, e.g., at densities close to the critical
density of a pion condensation, this would trigger efficient
direct Urca cooling processes. On the other hand, right from the
dispersion relation follows that\begin{align} \frac{m^*_{ \rm
F}}{m_{ \rm F}}=\frac{1-(\partial \Re\Sigma_{ \rm F}^R/\partial
\epsilon)_\rmF}{1+(\partial \Re\Sigma_{ \rm F}^R/\partial
\epsilon_p^0)_\rmF}\,,
\end{align}
where $\epsilon_p^0 =p^2/2m_{\rm F}$. Thereby, the Landau parameter
$\tilde{f}_1$ can be expressed via the energy-momentum derivatives
of the fermion self-energy.

Below we focus our study on effects associated with the zero harmonic
$\tilde{f}_0$ in the expansion of $\Gamma^{\om}_{0}$ as a function
of $(\vec{n}\vec{n}\,')$. Then the solution of Eq.~(\ref{Tph-FL})
is
\begin{eqnarray}
\widehat{T}_{\rm ph}(\vec{n}\,',\vec{n};q) &=& T_{{\rm ph},0}(q)\,
\sigma'_0\, \sigma_0 + T_{{\rm
ph},1}(q)\,(\vec{\sigma}\,'\vec{\sigma})\,, \nonumber\\ T_{{\rm
ph},0(1)}(q) &=& \frac{1}{1/\Gamma^\om_{0(1)}- \langle
\mathcal{L}_{\rm ph}(\vec{n};q)\rangle_{\vec{n}}}\,.
\label{Tph-sol}
\end{eqnarray}
The averaged particle-hole propagator is expressed through the Lindhard's function
\begin{align}
\langle \mathcal{L}_{\rm ph}(\vec{n};q)\rangle_{\vec{n}}=-a^2 N
\Phi\big(\frac{\om}{v_\rmF\,k},\frac{k}{p_\rmF}\big)\,,
\end{align}
where
\begin{align}
\Phi(s,x)&= \frac{z_-^2-1}{4(z_+-z_-)}\ln\frac{z_- + 1}{z_- - 1}
\nonumber\\ & -\frac{z_+^2-1}{4(z_+-z_-)}\ln\frac{z_+ + 1}{z_+ -
1} +\frac12\,. \label{LindF}
\end{align}
Here and below we use the dimensionless variables $z_\pm= s \pm x/2$, $s=\om/kv_{\rm F}$, $x=k/p_{\rm F}$. For real $s$ the Lindhard's function acquires an imaginary part
\begin{align}
\label{ImPhi}
\Im \Phi(s,x) =\left\{
\begin{array}{ccl}
 \frac{\pi}{2}\,s &,& 0\leq s\leq 1-\frac{x}{2}
\\
\frac{\pi}{4x}(1-z_{-}^2) &,&
1-\frac{x}{2}\leq  s\leq  1+\frac{x}{2}
\\
0 &,& \mbox{otherwise}
\end{array}
\right.
\,.
\end{align}
For $s \gg 1$,
\begin{align}
\Phi(s,x) \approx -1/(3\, z_+\, z_-)\,.
\end{align}
For $x\ll 1$ the function $\Phi$ can be expanded as
\begin{align}
\label{PhiLow-x}
\Phi (s, x)\approx
1 + \frac{s}{2} \log\frac{s-1}{s+1} -\frac{x^2}{12 \left(s^2-1\right)^2}\,,
\end{align}
and if we expand it further for $s\ll 1$ we get
\begin{align}
\label{PhiLow}
\Phi (s, x)\approx
1  +i \frac{\pi}{2}s -s^2 -\frac{x^2}{12} -\frac{s^4}{3}  -\frac{x^2s^2}{6} -\frac{x^4}{240}\,.
\end{align}

At finite temperatures the function $\Phi$ should be replaced by the
temperature dependent Lindhard function $\Phi_T$ calculated in Ref.~\cite{Voskresensky:1982vd}. Generalization of expansion~(\ref{PhiLow}) for low temperature case ($T\ll \epsilon_{\rm F}$) is given by
\begin{align}
\label{PhiT}
\Phi_T (s, x, t)=  \Phi(s,x)\, \Big(1-\frac{\pi^2}{12} t^2\Big)\,,
\end{align}
where $t=T/\epsilon_{\rm F}$ and $\epsilon_{\rm F}=p_{\rm F}^2/2m_{\rm F}^{*}$ is the Fermi energy. For high temperatures $(T\gg\epsilon_{\rm F}$) we have
\begin{align}
\label{PhiT1}
\Phi_T (0,x,t) =\frac{2}{3\, t}
\Big( 1-\frac{x^2}{6\,t}-\frac{1}{3\,\sqrt{2\pi\,t^3}}\Big)\,.
\end{align}

The amplitudes (\ref{Tph-sol}) possess simple poles and
logarithmic cuts. For the amplitude $T_{{\rm ph},0}$ the pole is
determined by the equation
\begin{eqnarray}
\frac{1}{f_0}
=-\Phi\big(\frac{\om}{v_\rmF\,k},\frac{k}{p_\rmF}\big)\,.
\label{sound-eq}
\end{eqnarray}
Similar equation exists in $g$-channel (with replacement $f_0\to
g_0$. Analytical properties of the solution (\ref{sound-eq}) have
been studied in \cite{Sadovnikova}.

 Expanding the retarded particle-hole amplitude $T^R_{{\rm
ph},0}(q)$ near the spectrum branch
\begin{align}
T^R_{{\rm ph},0}(q)&\approx \frac{2\om (k)V^2(k)}{(\om+i0)^2
-\om^2 (k)}\,, \nonumber \\ V^{-2}(k) & =a^2 N
\frac{\partial\Phi}{\partial \om^2}\Big|_{\om (k)}\,, \label{Tnr}
\end{align}
we identify the quantity $D^R(\om,k)=[(\om+i0)^2 -\om^2 (k)]^{-1}$
as the retarded propagator of a  boson with the dispersion
relation $\om =\om(k)$ and the quantity $V(k)$ as the effective
vertex of the fermion-boson interaction.

For the neutron matter  the parameters $f=f_{nn}$, $g=g_{nn}$ are
the neutron-neutron Landau scalar and spin parameters.
Generalization to the two component system, e.g., to the nuclear
matter of arbitrary isotopic composition is formally simple
\cite{M67}. Then the amplitude should be provided  with four
indices ($nn$, $pp$, $np$ and $pn$). However, equations for the
partial amplitudes do not decouple. For the isospin-symmetric
nuclear  system with the omitted Coulomb interaction the situation
is simplified since then $f_{nn}=f_{pp}$ and $f_{np}=f_{pn}$. In
this case one usually presents $\Gamma_0^\om$ in
Eqs.~(\ref{Gom-fullspin}) and (\ref{L-param}) as $\Gamma_0^\om
=(\tilde{f}+\tilde{f}'\vec{\tau}'\vec{\tau})/a^2N_0$, and
$\Gamma_1^\om
=(\tilde{g}+\tilde{g}'\vec{\tau}'\vec{\tau})/a^2N_0$, where
$\tau_i$ are isospin Pauli matrices. Quantities $f$ and $f'$ (and
similarly $g$ and $g'$) are expressed through $f_{nn}$ and
$f_{np}$ as $f=\frac{1}{2}(f_{nn}+f_{np})$ and
$f'=\frac{1}{2}(f_{nn}-f_{np})$.  The amplitudes of these four
channels ($f$, $f'$, $g$, $g'$) decouple, cf. Ref.~\cite{AAB}. The
excitation modes are determined by four equations similar to Eq.
(\ref{sound-eq}), now with $2f_0$, $2f_0'$, $2g_0$ and $2g_0'$.
Presence of the small Coulomb potential modifies the low-lying
modes determined by Eq. (\ref{sound-eq}) (now with $f_0$, $f_0'$,
$g_0$ and $g_0'$) only at low momenta \cite{AAB}. We will ignore
the Coulomb effects in our exploratory study.


We discuss now the properties of the bosonic modes for different values of the Landau parameter $f_0$.

\subsection{Repulsive interaction ($f_0>0$). Zero sound.}\label{sec:spec-f0pos}

For $f_0>0$  there exists a real solution of Eq.~(\ref{sound-eq})
such that for $k\to 0$ the ratio $\om_{s}(k)/k$ tends to a
constant. Such a solution is called a zero sound. The zero sound
exists as a quasi-particle mode in the high frequency limit $\om
\gg 1/\tau_{\rm col}$, where $\tau_{\rm col}\propto\epsilon_{\rm
F}/T^2$ is the fermion collision time. In the opposite limit it
turns into a hydrodynamic (first)
sound~\cite{Pethick-Ravenhall88}. At some value $k_{\rm lim}\lsim
p_\rmF$ the spectrum branch enters the region with $\Im\Phi> 0$,
and  the zero sound becomes damped diffusion  mode.

We search the solution of Eq.~(\ref{sound-eq}) in the form
\begin{align}
\om_s(k)=k\, v_\rmF\, s(x),
\end{align}
where $s(x)$ is a function of
$x=k/p_\rmF$ which we take in the form $s(x)\approx s_0 + s_2\,
x^2+s_4\,x^4$. The odd powers of $x$ are absent since $\Re\Phi$ is
an even function of $x$.

Above  we assumed that $f$ depends only on $\vec{n}\vec{n}'$. The
results are however also valid if the Landau parameter $f$ is a
very smooth function of $x^2$. From now we suppose
\begin{align}
\label{fexp}
f_0 (x)\approx f_{00} +f_{02} x^2\,,
\end{align}
where the parameter $f_{02}$ is determined by the effective range
of the fermion-fermion scattering amplitude and expansion is valid
provided $|f_{00}|\gg |f_{02}|$ for relevant values $x\lsim p_{\rm
F}$. According to Ref.~\cite{SaperFayans}, in the case of atomic
nuclei ($n\simeq n_0$), $f_{02}=- f_{00} r_{\rm eff}^2 p_{\rm
F}^2/2$, and  $0.5\lsim r_{\rm eff}\lsim 1$fm, as follows from the
comparison with the Skyrme parametrization of the nucleon-nucleon
interaction and with the experimental data. For isospin-symmetric
nuclear matter $f_{00}>0$ for $n\gsim n_0$, $f_{00}<0$ for lower
densities, and in a certain  density interval below $n_0$,
$f_0<-1$, cf. Refs.~\cite{Speth:2014tja,SVB}. In the purely neutron
matter one has $-1< f_{00} < 0$ for $n\lsim
n_0$~\cite{Wambach:1992ik}.

The constant term, $s_0$,  follows from Eq.~(\ref{sound-eq})
\begin{align}
\frac{1+f_{00}}{f_{00}}=\frac{s_0}{2}\ln\frac{s_0+1}{s_0-1}
\,. \label{s0-coeff}
\end{align}
For  $f_{00}\ll 1$ the solution of Eq.~(\ref{s0-coeff}) is
\begin{align}
s_{0}= & 1 + C \big[1 + (4+ 5 f_{00})(C/2 f_{00})
\nonumber\\ &+ (24 + 52 f_{00} + 29 f_{00}^{2})(C/2
 f_{00})^2\big]\, \,,
\end{align}
where $C=(2/e^2)\, \exp(-2/f_{00})$. For $f_{00}\le 1.8$ this expression reproduces the numerical solution with a percent precision. In the opposite limit $f_{00}\gg 1$ the asymptotic solution is $s_0 =\sqrt{f_{00}/3}$.
The  coefficients $s_2$ and $s_4$ follow as
\begin{align}
s_2 & = \left[\frac{f_{02}}{f_{00}^{2}} - \frac12 \frac{\partial^2
\Phi}{\partial x^2}\Big|_{s_0,0}\right] \Big[\frac{\partial
\Phi}{\partial s}\Big|_{s_0,0}\Big]^{-1} \label{s2-coeff}\\& =
\frac{ s_0 (\alpha +f_{02}) (s_0^2-1)}{f_{00}\,(1+ f_{00} -
s_0^2)}\,,\quad\alpha=f_{00}^2/[12(s_0^2-1)^2]\,. \nonumber
\end{align}
\begin{align}
s_4 &= s_0 f_{00}\frac{s_2^2 +
\frac{1 + 5 s_0^2-80(s_0^2-1) s_0 s_2}{240(s_0^2-1)^2}-\frac{f_{02}^2 }{f_{00}^{3}} (s_0^2-1)^2 } {(s_0^2-1) (1+f_{00}-s_0^2)}\,.
\label{s4-coeff}
\end{align}

\begin{figure}
\centerline{\includegraphics[width=6cm]{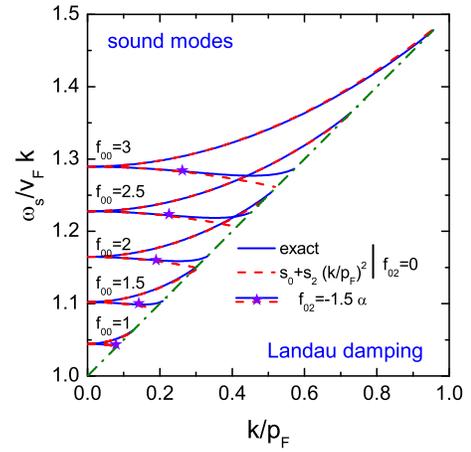}}
\caption{Spectrum of the sound modes defined by
Eq.~(\ref{sound-eq}) for various values of the Landau parameters
$f_{00}$ and $f_{02}$. Solid lines show the results of the
numerical solution, dashed lines are the quadratic expansions with
the parameters $s_0$ and $s_2$ given by Eqs.~(\ref{s0-coeff}) and
(\ref{s2-coeff}). The values of $f_{00}$ are shown in labels.
Lines with (without) stars are calculated for $f_{02}=0$
($f_{02}=-1.5\alpha$), where $\alpha$ is defined in Eq.
(\ref{s2-coeff}). The dash-dotted line shows the border of the
imaginary part of the Lindhard's function (\ref{LindF}). Below
this line the frequency is complex.} \label{fig:omS}
\end{figure}

The numerical solutions of Eq.~(\ref{sound-eq}) are shown in
Fig.~\ref{fig:omS}  by solid lines for various values of the
Landau parameters $f_{00}$ and $f_{02}$.  The zero-sound
solution exists for $s_0>0$.  We check that the inequality
$1+f_{00}-s_0^2
>0$ holds for $f_{00}>0$.  The quadratic approximation for the spectrum  $\om\approx
v_\rmF\,k\,[s_0+s_2 x^2]$ is demonstrated by dashed lines.
 Only the quasi-particle part of the spectrum   is
shown with $\om_s(k)/(v_\rmF\, k)>  1+ k/(2p_\rmF)$. This
inequality holds for $k/p_{\rmF}<k_{\rm
max}/p_\rmF=(1-\sqrt{1-16s_2(s_0-1)})/(4\,s_2)$ if $s_2\le
1/[16(s_0-1)]$. For larger $k$ the Lindhard function becomes
complex for continuation of the branch $\om=\om_s(k)$ and the
solution of Eq.~(\ref{sound-eq}) acquires an imaginary part. For
$s_2>1/[16(s_0-1)]$,   the function $\om_s(k)$ does not enter in
the region of the complex Lindhard function. However, for positive
$s_2$ there is another source of the mode dissipation related to
the decay of one mode's quantum in two quanta. The latter process
is allowed if the energy-momentum relation $\om_s (p) =\om_s
(p')+\om_s (|\vec{p}-\vec{p}{\,'}|)$ holds, that is equivalent to
the relation $ 1-\cos\phi  = 3s_2/(s_0 p_{\rm F}^2)$\,, where
$\cos\phi =(\vec{p}\,\vec{p}{\,'})/(p\,p')$, which is fulfilled if
$s_2>0$.

In Fig.~\ref{fig:omS} we demonstrate first the case $f_0={\rm
const}$, i.e., $f_{02}=0$. As we see, in this case $s_2>0$ for any
$f_{00}>0$. We see that the quadratic approximation coincides very
well with the full solution. Then we study how the spectrum
changes if the parameter $f_{02}$ is taken nonzero. As we conclude
from Eq.~(\ref{s2-coeff}) the coefficient $s_2$ can be negative
for $f_{02}<-\alpha\,.$ The latter implies: $f_{02}\lsim -10$ for
$f_{00}=1$; $f_{02}\lsim -2.6$ for $f_{00}=2$; and $f_{02}\lsim
-1.7$ for $f_{00}=3$. In Fig.~\ref{fig:omS}  by solid curves and
dashed curves marked with stars we depict the zero-sound spectrum
for $f_{02}< -\alpha$, being  computed following
Eq.~(\ref{sound-eq})  and within the quadratic approximation,
respectively. We chose here $f_{02}=-1.5 \alpha$. The quadratic
approximation for $s(x)$ works now worse for momenta close to
$k_{\rm max}$ and the next term $s_4 x^4$ should be included. We
note that the parameter $s_4$ is positive in this case and the
function $s(x)$ has a minimum at $x_{\rm min}=\sqrt{|s_2|/(2
s_4)}$. In the point $k=k_{0}$ corresponding to the minimum of
$\om_s(k)/k$ the group velocity of the excitation $v_{\rm
gr}=d\om_s/dk$ coincides with the phase one $v_{\rm ph}=\om_s/k$.
The quantity $\om_s (k_0)/k_0$ coincides with the value of the
Landau critical velocity $u_{\rm L}$ for the production of Bose
excitations in the  superfluid moving with the velocity $u>u_{\rm
L}$.

\begin{figure}
\centerline{\includegraphics[width=6cm]{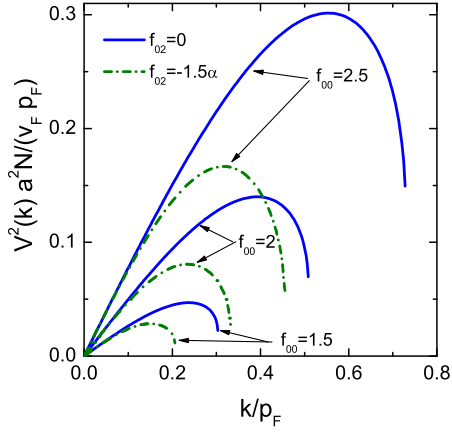}}
\caption{Coupling constant of zero-sound modes and
fermions as a function of the momentum $k$ for various values of
the Landau parameters $f_{00}$ and $f_{02}$, calculated from
Eq.~(\ref{Tnr}). All curves end at the limiting value of $x_{\rm
max}=k_{\rm max}/p_\rmF$ at which the spectral branch enters into
the region of complexity of the $\Phi$ function. } \label{fig:g2k}
\end{figure}

The effective vertex of the boson-fermion coupling (\ref{Tnr}) is
shown in Fig.~\ref{fig:g2k} as a function of $x$ for several
values of $f_{00}$ and $f_{02}$. We plot it in the range
$0<x<x_{\rm max}=k_{\rm max}/p_\rmF$, where the quasiparticle
branch of the zero-sound spectrum is defined. For $f_{02}=0$ the
vertex $V(k)$ is shown by solid lines, for $f_{02}< 0$ by
dash-dotted lines. We see that $V^2 (x_{\rm max})$ is smaller for
$f_{02}< 0$ than for   $f_{02}=0$. For small $x$  one can use the
analytical expression
\begin{align}
&a^2 N \frac{V^2(x)}{v_\rmF\, p_\rmF} =
x\Big[\frac{\partial\Phi}{\partial s}\Big|_{s(x),x}\Big]^{-1}
 \nonumber\\
&\quad \approx \frac{s_2 xf_{00}^2}{\alpha+f_{02}}\Big[1+\frac{4\alpha s_2 x^2
}{\alpha+f_{02}}\Big(6  s_2-\frac{s_0}{s_0^2-1}\Big)\Big]\,,
\label{g2k-exp}
\end{align}
which works well for $k\lsim 0.75 k_{\rm max}$. For $k\to 0$ we
get $a^2 N {V^2(k)}\simeq  s_0 (s_0^2-1)^2 f_{00}k
v_{\rmF}/(1+f_{00}-s_0^2)$.

\subsection{Moderate attraction, $-1<f_{00}<0$. Diffusons.}

Assume $f_{02}=0$. For $-1 < f_{00} < 0$, Eq.~(\ref{sound-eq}) has
only damped solutions with $\Re s<1 $ and $\Im s <0$.  In the
limit of $s,x\ll 1$, using the expansion (\ref{PhiLow}) we easily
find the analytic solution
\begin{align}
\label{omnegf}
\frac{\om_{\rm d}(k)}{kv_{\rm F}} =i s_{\rm d}(x) \approx -i\frac{2}{\pi} \left[\frac{1-|f_{00}|}{|f_{00}|}+\frac{x^2}{12}\right]\,,
\end{align}
which is valid for $1-|f_{00}|\ll 1$ and $x\ll 1$. We see that the
solution we found is purely imaginary and its dependence on $x$ is
very weak.

\begin{figure}
\centerline{\includegraphics[width=6cm]{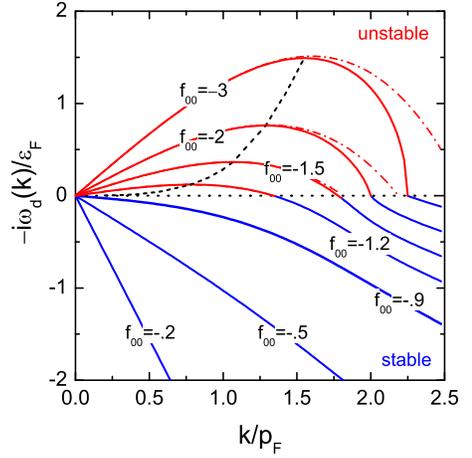}}
\caption{Solid lines show the purely imaginary solutions of
Eq.~(\ref{eq-omD}) for various values of the Landau parameter
$f_{00} <0$. Dashed line indicates positions of the maxima of
spectra $\om_{\rm d}$ for $f_{00}<-1$. Dash-dotted lines show
the approximated spectra $\om_{\rm d}(k)\approx i v_\rmF k (s_{\rm
d0}+ s_{\rm d2}\,x^2+ s_{\rm d4} x^4)$ with the parameters from
Eqs.~(\ref{sd0}) and (\ref{sd24}). } \label{fig:omD}
\end{figure}

More generally, for purely imaginary $s=i\,\tilde{s}$,
$\tilde{s}\in R$, the Lindhard function can be rewritten as
\begin{align}
&\Phi(i\tilde{s},x) =\widetilde{\Phi}(\tilde{s},x)= \frac{1}{4x}\big[\tilde{s}^2-\quart
x^2+1\big] \log \frac{\tilde{s}^2+\left(\half
x+1\right)^2}{\tilde{s}^2+\left(\half x-1\right)^2}
\nonumber\\
&\quad + \frac{{\tilde{s}}}{2}\left[{\rm
arctan}\frac{\tilde{s}}{1-\half x} + {\rm
arctan}\frac{\tilde{s}}{1+\half
x}\right]+\frac{1-\pi\tilde{s}}{2}\,.
\label{Phist}
\end{align}
We note that the function ${\rm
arctan}(\tilde{s}/[1-\half x])$ should be continuously extended for $x>2$ as
${\rm arctan}(\tilde{s}/[1-\half x])+\pi{\rm sign}\tilde{s}$\,.
Solutions of equation
\begin{align}
1/|f_{00}|=\Phi(i\tilde{s},x)
\label{eq-omD}
\end{align}
for $0>f_{00}>-1$ are depicted in Fig.~\ref{fig:omD} for
$f_{00}=-0.2; -0.5; -0.9$. These solutions are damped ($-i\om
<0$). For $f_{02}\neq 0$ (at the condition $0>f_0 (k)>-1$),   the
damped character of the solutions does not change.

\subsection{
Strong attraction, $f_{00}<-1$. Pomeranchuk
instability.}\label{sec:pominstab}

Now we consider the case of a strong attraction in the scalar
channel $f_{00}<-1$. We continue to assume $f_{02}=0$ for
certainty.  The Pomeranchuk instability for $f_{00}<-1$ manifests
itself in the negative compressibility which is found from the
relation for the variation of the fermion contribution to the
chemical potential~\cite{LP1981}
\begin{align}
 K=\frac{p^2_{\rmF}(n_0)}{3m_{\rmF}^*(n_0)}(1 + f_{00})\,,
\end{align}
being negative for $f_{00} <-1$.

Solutions of Eq.~(\ref{eq-omD}) for $f_{00}<-1$  are  shown in
Fig.~\ref{fig:omD}. We see that there is an interval of $k$, $0 <
k < k_{\rm d}$, where $-i\om_{\rm d}>0$ and, hence, the mode is
exponentially growing with time. This corresponds to the
Pomeranchuk instability in a Fermi system with strong scalar
attraction, see Ref.~\cite{Nozieres}. For $k>k_{\rm d}$ we get
$-i\om_{\rm d}(k>k_{\rm d})<0$, and the mode becomes stable again.
The  momentum $k_{\rm d}$ is determined from Eq.~(\ref{eq-omD})
for $\tilde{s}=0$:
\begin{align}
\frac{1}{f_{00}}=\frac{x_{\rm d}^2-4}{8x_{\rm d}} \log
\left|\frac{x_{\rm d} + 2}{x_{\rm d} - 2} \right|- \frac{1}{2}\,,\label{omD-end}
\end{align}
$x_{\rm d}=k_{\rm d}/p_\rmF$.
The instability increment $-i\om_{\rm d}$ has a maximum at some momentum $k_m$. The locations of maxima for different $f_{00}$ values are connected by dashed line in Fig.~\ref{fig:omD}.

The spectrum of the unstable mode can be written for $x<<1$ as
$s_{\rm d}(x)\approx s_{\rm d 0}+ s_{\rm d 2}x^2+s_{\rm d
4}\,x^4$, where the leading term is determined by the equation
\begin{align}
\frac{\pi}{2}-\arctan s_{\rm d 0}=\frac{z_f}{s_{\rm d0}}\,,\quad z_f=1-1/|f_{00}| \,.
\label{sd0}
\end{align}
The subleading terms are equal to
\begin{align}
s_{\rm d 2}&=\frac{s_{\rm d0}}{12(1+s_{\rm d0}^2)(z_f\, (1+s_{\rm
d0}^2) -s_{\rm d0}^2)}\,, \nonumber\\ s_{\rm d 4}&=\frac{1-5
s_{\rm d0}^2}{20(1+s_{\rm d0}^2)^2}s_{\rm d2} - \frac{4 s_{\rm d0}
s_{\rm d2}^2}{1+s_{\rm d 0}^2} -12 s_{\rm d2}^2\,.
\label{sd24}\end{align} This approximation of the spectrum is
illustrated in Fig.~\ref{fig:omD} by dash-dotted lines. The
approximation works very well for $-1.5<f_{00}<-1$ (dash-dotted
lines and solid lines coincide) {\it but becomes worse for smaller
$f_{00}$.}

For a slightly subcritical case, where $0<z_f\ll 1$, we obtain
\begin{align}
\label{omD-unst}
\om_{\rm d}(k) \approx i\frac{2}{\pi}v_{\rm F}k\,[z_f-k^2/(12p_\rmF^2)]\,.
\end{align}
The function $-i\om_{\rm d}$ has a maximum at $k_m =2p_{\rm F}
\sqrt{z_f}$ equal to
 \be\label{maxsound}
 -i\om_m=  ({8}/{3\pi}) v_{\rm F}p_{\rm F}\,
z_f^{3/2}\,.
 \ee


\subsection{Spinodal instability.}

For $f_{00}<-1$ not only the compressibility is negative but also
the square of the first sound velocity ~\cite{LP1981}
\begin{align}\label{u}
u^2 =\frac{\partial P}{\partial \rho}=\frac{p_{\rm F}^2}{3m_{\rm
F} m^{*}_{\rm  F}}(1+ f_{00})<0\,.
\end{align}
Here $P$ denotes the pressure and $\rho$ is the mass density. Note
that in the limit $T\to 0$ the isothermal and adiabatic
compressibilities coincide, whereas for $T\neq 0$ the difference
becomes substantial, see Ref.~\cite{VS:2010gf}. Also note that the
first sound exists in the hydrodynamical (collisional) regime,
i.e. for $\om\ll \tau_{\rm col}^{-1}\propto T^2$,  which is the
opposite limit  to the collision-less regime of the zero sound,
i.e. $\om\gg \tau_{\rm col}^{-1}$.

For the van der Waals equation of state the compressibility and
the square of the first sound velocity  prove to be negative in
the spinodal region. Thereby excitation spectrum  is unstable and
in simplest case of ideal hydrodynamics the growing mode is as
follows~\cite{VS:2010gf}
\begin{align}\label{hydromode}
 -i\om =k\sqrt{|u^2 -ck^2|}\,,
\end{align}
where $c$ is a coefficient associated with the surface tension of
the droplets of one phase in the other one.  Note that the viscosity
and thermal conductivity may delay formation of the hydrodynamic
modes. Maximum  in $k$ yields
\begin{align}\label{spinod}
-i\om_m =v_{\rm F}p_{\rm F}(|f_{00}|-1)/(6m_{\rm F}\sqrt{c})\,.
\end{align}

 The rate of the growing of the spinodal mode
decreases with increase of the surface tension of the droplets,
whereas the growing rate of the collision-less mode does not
depend on the surface tension. For
 \be\label{surftens}
c>\frac{\pi^2|f_{00}|^{3/2}}{256 m_{\rm F}^2 (|f_{00}|-1)}
 \ee
the zero-sound-like excitations (\ref{omD-unst}) would grow more
rapidly than excitations of the ideal hydrodynamic mode
(\ref{spinod}).

For isospin symmetric nuclear matter the spinodal region in the
dependence $P(1/n)$ exists at nucleon densities below the
saturation nuclear density, see Ref.~\cite{SVB}. In  multi
component systems with charged constituents, like neutron stars,
the resulting stationary state is the mixed pasta state, where
finite size effects (surface tension and the charge screening) are
very important, cf. Refs.~\cite{Maruyama:2005vb,Tatsumi:2002dq},
contrary to the case of the one component system, where the
stationary state is determined by the Maxwell construction, cf.
Ref.~\cite{SVB}. The liquid-gas phase transition may occur in
heavy-ion collisions. A nuclear fireball prepared in a course of
collision has a rather small size, typically less or of the order
of  the Debye screening length.  Therefore, the pasta phase is not
formed, as in a one-component system.

\section{Bosonization of the local interaction}\label{sec:boson}

The description of a fermionic system with a contact interaction
can be equivalently described in terms of bosonic fields
$\phi_{q}=\sum_{p} \psi^\dag_{p}\psi_{p+q}$, here we denote
$q=(\om,\vec{k})$ and $p=(\epsilon,\vec{p})$. In terms of the
functional path integral the transition to the collective bosonic
fields can be performed with the help of a formal change of
variables by means of a  Hubbard-Stratonovich
transformation~\cite{Altland-Simons,Kopietz}. After this
transformation the effective Euclidean action for the system with
a repulsive interaction, here $f_0 >0$, can be written within the
Matzubara technique in terms of the bosonic fields as
\begin{align}
S[\phi]=\sum_{q} \frac{\phi_{q}\, \phi_{-q}}{2\Gamma_0^\om} - \Tr
\log \Big[1 - \frac{\hat{G}i\hat{\phi}_q}{\beta}\,\Big]\,,
\label{effact}
\end{align}
where $\hat{G}$ and  $\hat{\phi}$ are infinite matrices in
frequency/momentum space with matrix elements $[\hat
G]_{p,p'}=\delta_{p,p'} G(\epsilon,\vec{p}\,)$ and
$[\hat{\phi}]_{q,q'}=\delta_{q,q'} \phi_{\vec{q}}$. Trace is taken
over frequencies and momenta and includes factor 2 accounting for
the fermion spins. Transformation to zero temperature follows by
the standard replacement $\beta^{-1}\sum_n \to\int d\epsilon/(2\pi
i)$.

Expanding Eq.~(\ref{effact}) up to the 4-th order in $\phi$ for
$T=0$ we obtain
\begin{align}
&S[\phi] \approx\frac12 \mbox{sgn} (f_0) \sum_{q}
\phi_{q}\,\big((\Gamma_0^\om)^{-1} + a^2N\Phi(s,x)\big)\phi_{-q}
\label{effact4}\\ &\qquad +\frac14\sum_{\{q_i\}}U(q_1,q_2,q_3,q_4)
\phi_{q_1}\phi_{q_2}\phi_{q_3}\phi_{q_4}\delta_{q_1+q_2+q_3+q_4,0}
\,,\nonumber
\end{align}
where the effective field self--interaction is given by the
function
\begin{align}
U(q_1,q_2,q_3,q_4)
&=-2i\frac{1}{4!}\sum_{\mathcal{P}(1,\dots,4)}\sum_p
G_{p}G_{p+q_1} \nonumber\\ &\times
G_{p+q_1+q_2}G_{p+q_1+q_2+q_3}\,. \label{Lambda-gen}
\end{align}
Here the sum $\sum_{\mathcal{P}}$ runs over the $4\!$ permutations
of $4$ momenta $q_i$. As we will show, Eq. (\ref{effact4}) is
applicable also for $f_0<0$ after the replacement $\phi\to i\phi$
 that results in the appearance of the prefactor $\mbox{sgn} (f_0)$ in Eq.
(\ref{effact4}).

The general analysis of Eq.~(\ref{effact4})  for
arbitrary external momenta was undertaken in Ref.~\cite{Brovman}.
The first term in Eq.~(\ref{effact4})  (quadratic in $\phi$) can
be interpreted as the inversed retarded propagator of the
effective boson zero-sound-like mode:
\begin{align}
\label{DS}
(D^R_{\phi})^{-1}(\om,k)= -{\rm sgn}(f_0)[(\Gamma_0^\om)^{-1} +
a^2N\Phi(s,x)]\,.
\end{align}

To describe modes for an attractive interaction, $f_0<0$, it is
instructive to re-derive the expression for the mode propagator
using the approach proposed in Ref.~\cite{IKV00}. The local
four-fermion interaction $\Gamma_0^\om$  can be viewed as an
interaction induced by the exchange of  a scalar heavy boson with
the mass $m_{\rm B}$,
\begin{align}
\label{Gamom} \Gamma_0^\om \to -\frac{\Gamma_0^\om m_{\rm
B}^2}{(\om+i0)^2-k^2-m_B^2}\equiv -\Gamma_0^\om\, m_{\rm
B}^2\,D^R_{\rm B,0}(\om,k)\,,
\end{align}
where $D^R_{\rm B,0}$ is the bare retarded Green's function of the
 heavy boson and we assumed that $m_{\rm B}^2$ is much larger
then typical squared frequencies and momenta, $\om^2, k^2$, in the
problem.

Then making the replacement (\ref{Gamom}) in  the fermion
scattering amplitude (\ref{Tph-sol}) we obtain
\begin{align}
T^R_{{\rm ph},0}(\om,k)&= \frac{-(a^2N)^{-1}f_0 m_{\rm
B}^2}{(D_{{\rm B}, 0}^R)^{-1}(\om,k) - \Sigma^R_{\rm B}(\om,k)  }
 \equiv  V_{\rm B}^2 D^{\rm R}_{\rm B}(\om,k)\,.
\label{Tph-sol1}
\end{align}
Hence, we can identify the vertex of the boson-fermion interaction
\begin{align}\label{vert}
V_{\rm B}^2= - f_0 m_{\rm B}^2/(a^2N)\,,
\end{align}
and the full retarded Green's function of the boson $D^R_{\rm B}(\om,k)$ with the retarded self-energy
\begin{align}
\label{SigmaR}
\Sigma^R_{\rm B}(\om,k) = f_0 m_{\rm B}^2 \Phi\big(\frac{\om}{kv_\rmF},\frac{k}{p_\rmF}\big)\,.
\end{align}
Note that for very large $m_{\rm B}$ that we have assumed, $D^{\rm
R}_{\rm B}(\om,k)$ in (\ref{Tph-sol1}) differs from $D^{\rm
R}_{\phi}(\om,k)$ in (\ref{DS}) only by a prefactor. For the
attractive interaction, which we are now interested in ($f_{0}<0$),
we have $m_{\rm B}^2 >0$ and $V_{\rm B}^2>0$. For repulsive
interaction instead of the bare scalar boson the bare vector boson
would be an appropriate choice, cf. ~\cite{IKV00}, or we may come
back to the formalism given by Eqs. (\ref{effact}) --
(\ref{Lambda-gen}).

The spectrum follows from the solution of the Dyson equation $0=[D_{\rm
B}^R(\om,k)]^{-1}\approx -m_{\rm B}^2(1+f_{0}\Phi(s,x))$ for
$m_{\rm B}\gg \om,k$ and coincides with the solution of
Eq.~(\ref{sound-eq}).
%
The boson spectral function is given by
\begin{align}
&A_{\rm B}(\om,k)=-2\Im D_{\rm B}^R(\om,k), \nonumber\\ &-2\Im
T^R_{{\rm ph},0}(\om,k)=V_{\rm B}^2 A_{\rm B}(\om,k)\,.
\end{align}

\section{Avoiding of the Pomeranchuk instability by a Bose condensation}\label{sec:pomer}

\subsection{Condensation of a scalar field for $f_0<-1.$}

Consider a Fermi liquid with local scalar interaction $f_0<-1$. We
assume that the bosonization procedure of the interaction,
Eqs.~(\ref{Gamom}) and~(\ref{Tph-sol1}), is performed. According
to the perturbative analysis in Sect.~\ref{sec:pominstab}, for
$f_0<-1$ there are modes which grow with time. The growth of
hydrodynamic modes (first sound), cf. Eqs.~(\ref{hydromode}) and
(\ref{spinod}), results in the formation of a mixed phase. Besides the
hydrodynamic modes the zero-sound-like modes grow with time, cf. Eqs.~(\ref{omD-unst}) and (\ref{maxsound}). We study now the opportunity
that the instability of the zero-sound mode may result in a
formation of a static condensate of the scalar field. This leads
to a decrease of the system energy and to a rearrangement of the
excitation spectrum on the ground of the condensate field.

We will exploit  the simplest probing function describing the
complex scalar field of the form of a running wave
\begin{align}\label{running}
\varphi =\varphi_0 e^{-i\om_c  t+i\vec{k}_c\vec{r}}\,,
\end{align}
with the condensate frequency and momentum $(\om_c, \vec{k}_c)$
and the constant amplitude $\varphi_0$. The choice of the structure of
the order parameter is unimportant for our  study of the stability
of the system.

Guided by the construction of the full Green's function of the
effective boson field, see Eq.~(\ref{Tph-sol1}), the effective
Lagrangian density for the condensed field (\ref{running}) can be
written as, cf. Ref.~\cite{MSTV90,V84},
\begin{align}
\label{Lagr-B} L=\Re D_{\rm B}^{-1} (\om_c, k_c) |\varphi_0|^2
-{\textstyle\frac12}V_{\rm B}^4\Lambda(\om_c, k_c)
|\varphi_0|^4\,.
\end{align}
The energy density of the condensate is given then by the standard
relation
\begin{align}
E_{\rm B} =\om \partial L/\partial \om -L \,.
\nonumber
\end{align}
Fully equivalently, this Lagrangian density can be written in the
form suggested by expansion~(\ref{effact4}), now applied for
the running wave classical field, after the field redefinition
$m_{\rm B}\varphi_0=a\sqrt{N/|f_0|}\phi_0$,
\begin{align}
\label{Lagr} L = \Re D_{\phi}^{-1}(\om_c,k_c) |\phi_0|^2
 -{\textstyle\frac12}\Lambda(\om_c, k_c) |\phi_0|^4\,.
\end{align}
Here and in Eq.~(\ref{Lagr-B}) the quantity $\Lambda(\om_c, k_c)$
represents the self-interaction amplitude of condensed modes which
corresponds to the ring diagram with four fermion Green's
functions (\ref{Lambda-gen}),
$\Lambda(\om_c,k_c)=6U(q_c,-q_c,q_c,-q_c)$.  As we shall see below
the energetically favorable is the state with $\om_c =0$, and
therefore we put $\om_c=0$ here. The leading order contribution to
the self-interaction parameter $\Lambda(0,k_c)$ as a function of
the condensate momentum was calculated in Ref.~\cite{Brovman}:
\begin{align}
\Lambda(0,k_c)&=\frac{8 a^4}{\pi^2 v_\rmF^3 x_c^4}\Big[
\frac{1-x_c^2}{x_c}\ln\Big|\frac{2+x_c}{2-x_c}\Big| -
\frac{1-x_c^2}{2x_c}\ln\Big|\frac{1+x_c}{1-x_c}\Big| \nonumber\\
&+ \frac{x_c^2}{4-x_c^2}\Big]\,,\quad x_c=k_c/p_\rmF.
\end{align}
For $x_c\ll 1$ we get
\begin{align}
\Lambda(0,k_c)\approx
a^4\lambda\,\Big(1+\frac{k_c^2}{2p_\rmF^2}\Big)\,,\quad \lambda =
\frac{1}{\pi^2v_\rmF^3}\,. \label{lamb-c}
\end{align}
The quantity $\lambda$ agrees also with the result derived in
Ref.~\cite{D82} for description of the pion condensation in the
Thomas-Fermi approximation.

The equation of motion for the field amplitude follows from the
variation of the action corresponding to the Lagrangian density
(\ref{Lagr}),
\begin{align} \label{eq4fi0}
-a^2N\tilde{\om}^2(k_c)\phi_0 -\Lambda(0,k_c)\,|\phi_0|^2\phi_0 =
0\,,
\end{align}
where we introduce the effective boson gap
\begin{align}\label{omti}
\tilde{\om}^2(k_c)=[1-|f_0(k_c)|\Phi(0,k_c)]/|f_0(k_c)|\,,
\end{align}
 as it was done in the description of the pion condensation, see
Ref.~\cite{MSTV90}. The  equation for the condensate amplitude
(\ref{eq4fi0})  has a non trivial solution for
$\tilde{\om}^2(k_c)<0$,
\begin{align}
|\phi_0|^2 =-N\frac{\tilde{\om}^2(k_c)}{
 a^2\lambda\,\Big(1+\frac{k_c^2}{2p_\rmF^2}\Big)}\,\theta(-\tilde{\om}^2(k_c)) \,,
 \label{varphi0}
\end{align}
which corresponds to the gain in the energy density
\begin{align}\label{Eb}
E_{\rm B}=- N^2\frac{\tilde{\om}^4(k_c)}{ 2
\lambda\,\Big(1+\frac{k_c^2}{2p_\rmF^2}\Big)}\theta(-\tilde{\om}^2(k_c))\,,
\end{align}
where $\theta (x)=1$ for $x>0$ and 0 otherwise.

\subsection{Bose condensation in a homogeneous state.}

If the minimum of the gap $\tilde{\om}^2$ is realized at $k_c=0$,
e.g., it is so for $f_{02}\ge 0$, the energy density $E_{\rm B}$
is minimized for $\om_c=k_c=0$. Then, the field amplitude and the
condensate energy density are
\begin{align}
\label{varphi0-00} \phi_0^2 = N\frac{z_f}{a^2
\lambda}\,\theta(z_f)\,, \quad E_{\rm B}=-N^2\frac{z_f^2}{2
\lambda}\,\theta(z_f) \,,
\end{align}
where we used the same notation $z_f$ as in Eq.~(\ref{sd0}). The
Bose condensate is formed for $f_{00}<-1$.

For the case of non-zero temperature, $T\neq 0$, in the mean field
approximation one should just replace $\Phi\to \Phi_T$, where the
fermion step-function distributions are replaced to the thermal
distributions, and from Eq.~(\ref{PhiT}) for $T\ll \epsilon_{\rm
F}$ one recovers the critical temperature of the condensation
$T^{\rm MF}/\epsilon_\rmF=\sqrt{12\,z_f}/\pi$ valid for $0<z_f\ll
1$. For $|f_{00}|\gg 1$ from Eq.~(\ref{PhiT1}) one gets $T^{\rm
MF}/\epsilon_{\rm F} = 2|f_{00}|/3$\,.

In the presence of the homogeneous condensate ($k_c=0$) setting
$\phi_q =\phi_0+\phi'_q$ in (\ref{effact4}), (\ref{Lagr}), varying
in $\phi'_q$ and retaining only terms linear in $\phi'_q$ we
recover the spectrum of overcondensate excitations
\begin{align}
a^2 N\Big[-|f_{00}|^{-1} + \Phi\Big(\frac{\om}{v_\rmF
k},\frac{k}{p_\rmF}\Big)\Big]-\delta\Sigma_\varphi=0,
\label{dispeq-cond}\end{align} where the last term arises from the
interaction of excitations with the condensate.
Making use of Eqs~(\ref{PhiLow}), (\ref{lamb-c}), and
(\ref{varphi0-00}) we find
\begin{align}
\delta\Sigma_\varphi=2\,a^2N\,z_f +O(\om^2, k^2, \om/k\to 0)
\label{dSigfi}
\end{align}
and derive the spectrum for $z_f\ll 1$ and $x\ll 1$,
\begin{align}
\label{newspectr}
 \tilde{\om} \approx -i\frac{2}{\pi}z_f kv_{\rm F}\,.
\end{align}
So, the overcondensate excitations are damped, similar to those we
obtain for the case $-1<f_{00}<0$, see Eq.~(\ref{omnegf}).

In the presence of the condensate ($f_{00}<-1$) the particle-hole interaction is described by scattering amplitude
\begin{align}
&a^2\,N\,T_{\rm ph,0}^R =\Big[\frac{1}{f_0}+
\Phi-\frac{\delta\Sigma_\varphi}{a^2N} \Big]^{-1} \nonumber
\\&\simeq \Big[\frac{1}{f_{\rm ren, 0}}+ 1+i\frac{\pi\om}{2kv_{\rm
F}} +O(\om^2 ,k^2, \om/k\to 0)\Big]^{-1}\,,
\end{align}
where in the second equality we introduce the renormalized local
interaction
\begin{align}
&f_{\rm ren,0}\approx f_{\rm ren,00}
= -\frac{f_{00}}{2f_{00}+1}\,.
\label{fren}
\end{align}
The latter equation shows that if originally we have $f_{00}<-1$,
the renormalized interaction $-1<f_{\rm ren,00} < -1/2$. Thus, in the
presence of  the boson condensate the Fermi liquid is
free from the Pomeranchuk instability.

Using the renormalized parameter $f_{\rm ren,00}$ we can calculate how the fermion chemical potential changes in the presence of the condensate. We use here the standard expression
from Ref.~\cite{M67} for the variation of the chemical potential with the particle density
in the Fermi liquid theory
\begin{align}
\delta \mu_{\rm F} =\frac{2\epsilon_{{\rmF},0}}{3n_0} \delta n
+ \frac{f_{\rm ren,00}}{N}\delta n \,,
\end{align}
here $\epsilon_{{\rmF},0}=p_{\rm F}^2 (n_0)/2m^{*}_{\rm F} (n_0)$.
The energy density can be obtained from the relation
$E=\int_0^{\mu_{\rm F}} \mu_{\rm F} dn$.

The compressibility of the system becomes now
\begin{align}
K&=n_0\frac{\delta\mu_\rmF}{\delta n}\Big|_{n_0}+K_{\rm B}
\nonumber\\ &=\frac{p^2_{\rm F}(n_0)}{3m_{\rm
F}^*(n_0)}\big(1+f_{\rm ren,00}\big)+K_{\rm B}\,, \label{Kcond}
\end{align}
where the first term in the second line is always positive
\begin{align}
1+f_{\rm ren,00} = (|f_{00}|-1)/(2|f_{00}|-1)>0\,,
\end{align}
for $f_{00}<-1$ under consideration. The second term  is the
condensate term:
\begin{align}
K_{\rm B }=n_0\frac{d^2 E_{\rm B}}{d n^2} \Big|_{n_0}
\,.
\end{align}

Similarly, the square of the first sound velocity in the presence
of the condensate becomes
\begin{align}
u^2 =\frac{\partial P}{\partial \rho}=\frac{p_{\rm F}^2}{3m_{\rm F} m^{*}_{\rm F}}
(1+f_{\rm ren,00})>0\,,
\end{align}
whereas in the absence of the condensate it was $\propto
(1+f_{00})<0$, cf. Eq. (\ref{u}).

Recall that $u^2$ is
negative within the spinodal region which exists, if the pressure
has a van-der-Waals form. Thus, the formation of the scalar Bose
condensate, suggested here, might compete with the development of
the spinodal instability at the liquid-gas phase transition in
Fermi liquids. From general principles, one can expect that the
system will develop a (stable or metastable) condensate state
at least, if a surface tension between liquid and gas regions is
sufficiently high. Then aerosol-like mixture appearing in the
course of the development of the spinodal decomposition will
evolve more slowly compared to the process of the formation of the
scalar condensate.

\subsection{Bose condensation in an inhomogeneous state.}

\begin{figure}
\centerline{\includegraphics[width=6cm]{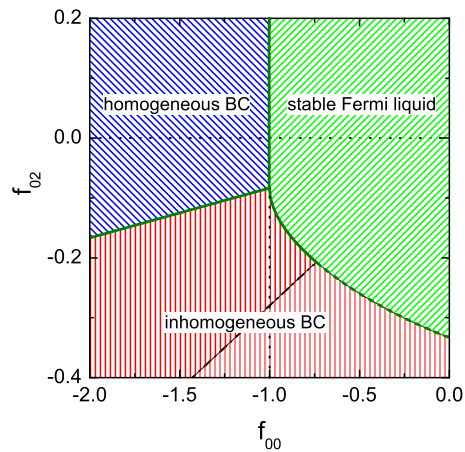} }
\caption{Phase diagram of a Fermi liquid with a scalar interaction
in the particle hole channel described by the momentum dependent
coupling constant $f_0(k/p_\rmF)=f_{00}+f_{02}k^2/p_\rmF^2$.
Depending on the values of parameters $f_{00}$ and $f_{02}$ the
Fermi liquid can be either in a stable state or in  states with
homogeneous or inhomogeneous static Bose condensates (BC) of
scalar modes. Possibility of the development of the spinodal
instability is suppressed.} \label{fig:PhaseD}
\end{figure}

We discuss now how the momentum dependence of the coupling
constant $f_0$, as given by Eq.~(\ref{fexp}), may influence the
condensation of the scalar mode. The  condition for the appearance
of a condensate with $\om_c=0$ and a finite momentum $k_c$ are
determined by relations $\tilde{\om}^2(k_c)\le 0\,,\quad
\frac{\rmd}{\rmd k}\tilde{\om}^2(k_c)=0$\,. The critical condition
for the appearance of the condensate can be deduced from the
expansion of $|f_0|\tilde{\om}^2(k)$ in (\ref{omti}) for $x^2\sim
x_m^2$:
\begin{align}
&1+f_0(x)\Phi(0,x)\approx 1 +f_{00} + 3\frac{
(f_{02}-\frac{f_{00}}{12})^2}{(f_{02} + \frac{f_{00}}{20})}
\nonumber\\
&-\frac{ f_{02} +\frac{f_{00}}{20}  }{12} [x^2 - x_m^2]^2\,,
\quad
x_m^2=6 \frac{f_{02}-\frac{f_{00}}{12}}{f_{02} + \frac{f_{00}}{20}}.
\end{align}
For $f_{00}<-1$ this function has always a negative minimum at
$x=0$, the second and deeper minimum appears at finite $x$, if
\begin{align}
f_{02}\le f_{02}^{\rm cr}={f_{00}}/{12}\,. \label{cond1}
\end{align}
For $-1<f_{00}<0$ the minimum at finite momentum can be realized if
\begin{align}
f_{02}\le  f_{02}^{\rm cr}=-\frac16\Big[1+\half
f_{00}+\sqrt{1+\frac{2}{5}f_{00}-\frac35 f_{00}^2}\Big]\,,
\label{cond2}
\end{align}
and
\begin{align}
x_m^2(f_{02}^{\rm
cr})=6\frac{1+f_{00}+\sqrt{1+\frac{2}{5}f_{00}-\frac35
f_{00}^2}}{1+\sqrt{1+\frac{2}{5}f_{00}-\frac35 f_{00}^2}+\frac15
f_{00}}\,.
\end{align}

 The solid lines in Fig.~\ref{fig:PhaseD} on the
$(f_{00},f_{02})$ plane show the  critical values  $f_{02}^{\rm
cr}$ given by Eqs. (\ref{cond1}) and (\ref{cond2}). They form the
phase diagram of the Fermi liquid with the momentum dependent
interaction in the scalar channel.  The dash-dotted line  shows
solution of equation $f_{02}x_m^2 =f_{00}$. In the region
restricted by this dash-dotted line and the dash continuation of
the solid line we have $|f_{00}|< |f_{02}|x_m^2 $ and the
expansion (\ref{fexp}) for $f_0 (k)$ becomes questionable.

Near the minimum the boson gap  $\tilde{\om}^2(k)$ can be
presented as
\begin{align}
\tilde{\om}^2 (k) =\tilde{\om}^2_0 (k_c) +\frac{\gamma}{4k_c^2}
(k^2 -k_c^2)^2
 \,,
 \label{gap-min-exp}
\end{align}
for $\gamma ={\rm const}$. The critical point is found from the
condition  $\tilde{\om}^2_0 (k_c=k_{\rm cr})=0$. Beyond the
critical point the actual value of $k_c$ follows from the minimum
of the energy density (\ref{Eb}) as a function of $k_c$.

\section{Condensate of Bose excitations  with non-zero momentum
in a moving Fermi liquid}\label{sec:moving}

 Let us apply the constructed above formalism to the analysis of
a possibility of condensation of zero-sound-like excitations with a
non-zero momentum and frequency in a moving Fermi liquid.  The
main idea was formulated in
Refs.~\cite{Pitaev84,V93,Vexp95,Melnikovsky,BP12,KV2015,Kolomeitsev:2015dua}.
When a medium moves in straight line with the velocity $u>u_{\rm
L}={\rm min}[\om (k)/k]$ (where the minimum is realized at
$k=k_0\neq 0$),  it may become energetically favorable to transfer
a part of the momentum from the particles of the moving medium to
a condensate of collective Bose excitations with the momentum
$k_0$. The condensation may occur, if in the spectrum branch $\om
(k)$ there is a region with a small energy at sufficiently large
momenta.

As in Ref.~\cite{V93}, we consider a fluid element of the medium
with the mass density $\rho$ moving with a non-relativistic
constant velocity $\vec{u}$. The quasiparticle energy $\om (k)$ in
the rest frame of the fluid is determined from the dispersion
relation
\begin{align}
\label{disp} \Re D_{\phi}^{-1}(\om, k)=0\,.
\end{align}
We continue to exploit the complex scalar condensate field
described by the simplest running-wave probing function, cf.
Eq.~(\ref{running}), and the Lagrangian density (\ref{Lagr}), but now
for the condensate of excitations.

The appearance of the condensate with a finite momentum
$\vec{k}_0$, frequency $\om =\om (k_0)$ and an amplitude
$\varphi_0$ leads to a change of the fluid velocity from $\vec{u}$
to $\vec{u}_{\rm fin}$, as it is required by the momentum
conservation
\begin{align}
\label{momentumcons} \rho \vec{u}
=\rho \vec{u}_{\rm fin} +\vec{k}_0 Z^{-1}_0 |\phi_0^2| \,,
\end{align}
where $\vec{k}_0 Z^{-1}_0 |\phi_0^2|$ is the density of the momentum
of the condensate of the boson quasiparticles with the
quasiparticle weight
\begin{align} Z^{-1}_0 (k_0) = \Big[
\frac{\partial}{\partial \om}\Re D^{-1}_{\phi}(\om,
k)\Big]_{\om(k_0),k_0}>0\,. \label{Zinv}
\end{align}

If in the absence of the condensate of excitations the energy
density of the liquid element was $E_{\rm in}=\rho u^2/2$, then in
the presence of the condensate of excitations, which takes a part
of the momentum, the energy density becomes
\begin{align}
\label{Ef} E_{\rm fin}=\half\rho u_{\rm fin}^2 +\om (k_0)Z^{-1}_0
|\phi_0|^2 +\half \Lambda (\om (k_0), k_0) |\phi_0|^4.
\end{align}
Here the last two terms appear because of the classical field of the condensate of  excitations.
The gain in the energy density due to the condensation, $\delta E=E_{\rm fin} -E_{\rm in}$, is equal to
\begin{align}
\delta E = -[\vec{u}\vec{k}_0 -\om(k_0)]\,Z^{-1}_0(k_0)\,
|\phi_0|^2 +\half\widetilde{\Lambda}|\phi_0|^4\,,
\end{align}
where
\begin{align}
\label{lambdatilde} \widetilde{\Lambda}=\Lambda (\om
(k_0),k_0)+(Z^{-1}_0(k_0))^2 k_0^2/\rho\,.
\end{align}
For $\om =0$, $\Lambda(0, k_0)$ is calculated explicitly, cf.
Eq.~(\ref{lamb-c}). Note that above equations hold also for
$\Lambda =0$.

The condensate of excitations is generated for the velocity of the
medium exceeding the Landau  critical velocity, $u>u_{\rm L} =\om
(k_0)/k_0$, where the direction of the condensate vector
$\vec{k}_0$ coincides with the direction of the velocity,
$\vec{k_0}\parallel \vec{u}$, and the magnitude $k_0$ is
determined by the equation $\om(k_0)/k_0 =\rmd\om(k_0)/\rmd k$.
The gain in the energy density after the formation of the
classical condensate field with the amplitude $\phi_0$ and the
momentum $k_0$ is then
\begin{align}\label{condE}
\delta E =- Z^{-1}_0(k_0) [u k_0 -\om (k_0)]\phi_0^2 + \half\widetilde{\Lambda} \phi_0^4\,.
\end{align}
The amplitude of the condensate field is found by minimization of the energy. From (\ref{condE}) one gets
\begin{align}
\label{varphi}
\phi_0^2 =Z^{-1}_0 (k_0)\frac{u k_0 -\om (k_0)}{\widetilde{\Lambda}}\theta (u-u_c)\,.
\end{align}
The resulting velocity of the medium becomes
\begin{align}
\label{finvel} u_{\rm fin} =u_c +\frac{(u-u_c)\theta
(u-u_c)}{1+[Z^{-1}_0 (k_0)]^2 k_0^2/(\Lambda (\om
(k_0),k_0)\rho})\,.
\end{align}
For a small $\Lambda$, we have $u_{\rm fin} =u_c +O(\Lambda)$.




For the repulsive interaction $f_0>0$ there is real zero-sound
branch of excitations $\omega_{\rm s} (k)\approx k v_\rmF (s_0 +
s_2\, x^2+s_4\,x^4)$, where the parameters $s_i$ depend on the
coupling constants $f_{00}$ and $f_{02}$ according to
Eqs.~(\ref{s0-coeff}), (\ref{s2-coeff}), and (\ref{s4-coeff}). As
shown in Sect.~\ref{sec:spec-f0pos} the ratio $\omega_{\rm s}
(k)/(k v_\rmF)$ has a minimum at $k_{0}=p_\rmF\sqrt{-s_2/(2s_4)}$
provided $f_{02}$ is smaller than $f_{\rm crit,
02}=-f_{00}^2/[12(s_0^2-1)^2]$.
The Landau critical velocity of the medium is  equal to $u_{\rm
L}/v_\rmF=\om_{\rm s}(k_0)/(v_\rmF k_0) \approx s_0-s_2^2/(2s_4)$.
The quasiparticle weight of the zero-sound mode~(\ref{Zinv}) is
\begin{align}
Z_0^{-1}(k_0)=\frac{a^2
N}{k_0v_\rmF}\frac{\partial\Phi(s,x)}{\partial
s}\Big|_{\frac{\om_s(k_0)}{v_\rmF k_0},\frac{k_0}{p_\rmF}}\,.
\label{Z-fact}
\end{align}
The amplitude of the condensate field~(\ref{varphi}) can be written as
\begin{align}
\label{phif} \varphi^2_0 =Z_0^{-1}(k_0) k_0\frac{u -u_{\rm L}
}{\widetilde{\Lambda}} \theta(u-u_{\rm L})\, .
\end{align}
  The energy density gained
owing to the condensation of the excitations is
\begin{align}\label{deltaEex}
\delta E=- k_0^2(Z^{-1}_0(k_0))^2\frac{(u -u_{\rm L}
)^2}{2\widetilde{\Lambda}}\theta
 (u-u_{\rm L})\,.
\end{align}
For a small $\lambda$, Eqs.~(\ref{phif}), (\ref{deltaEex}) simplify as
 \begin{align}
\varphi^2_0 &=\rho \frac{(u -u_{\rm L})}{Z^{-1}_0(k_0) k_0}\theta
(u-u_{\rm L}), \nonumber\\ \delta E &=-\frac{\rho}{2}(u-u_{\rm
L})^2 \theta (u-u_{\rm L})\,.
 \end{align}

\section{Concluding remarks}\label{sec:conclude}

 We described the spectrum of scalar excitations in normal Fermi
liquids for various values of the Landau parameter $f_0$ in the
particle-hole channel and for different models of its momentum
dependence. For $f_0 >0$ we found a condition on the momentum
dependence of $f_0(k)$ when the zero-sound excitations with a
non-zero momentum can be produced in the medium moving with the
velocity larger than the Landau critical velocity. Such
excitations will form an inhomogeneous Bose condensate.  For
$-1<f_0 <0$ there exist only damped excitations. For $f_0<-1$ we
studied the instability of the spectrum with respect to the growth
of the zero-sound-like excitations (Pomeranchuck instability) and
the excitations of the first sound. The surface tension
coefficient above which the zero-sound-like mode grows more
rapidly than the hydrodynamic one (for ideal hydrodynamics) is
estimated.  Then we derived an effective Lagrangian for the
zero-sound-like modes by performing bosonization of a local
fermion-fermion interaction. We argue that the  Fermi liquid with
$f_0 <-1$ might become stable owing to appearance  of the static
(homogeneous or inhomogeneous) Bose condensate of the scalar
quanta. Properties of the novel condensate state are investigated.

In the future it would be important to search for a possibility of
realization of this phenomenon in some Fermi liquids. It would be
interesting to perform a careful comparative study of the
possibilities of the Bose condensation and the ordinary spinodal
instability appearing at the first-order phase transitions in the
Fermi systems. For this, an explicit expression for the energy
density functional for a specific system, e.g. for the nuclear
matter, is required. Such a programme might be realized within the
relativistic mean-field model. Similarly, we expect a
stabilization of the Fermi liquid with  a strong attractive
spin-spin interaction $g_0 <-1$ by a condensate of a virtual boson
field at certain conditions. These questions will be considered
elsewhere.

\acknowledgments The work was supported  by the Ministry of
Education and Science of the Russian Federation (Basic part), by
Grants No. VEGA-1/0469/15 and No. APVV-0050-11, by
``NewCompStar'', COST Action MP1304, and by Polatom ESF network.



\begin{thebibliography}{99}

\bibitem{FL}
L.D.~Landau, Sov. Phys. JETP {\bf 3}, 920 (1956); {\bf 5}, 1011
(1957);  {\bf 8}, 70 (1959).

\bibitem{LP1981}
E.M.~Lifshitz and L.P.~Pitaevskii, {\it Statistical Physics,
Part 2} (Pergamon, 1980).

\bibitem{Nozieres}
D.~Pines and Ph.~Nozieres, {\it The Theory of
Quantum Fermi Liquids} (W.A. Benjamin, N.Y., 1966).

\bibitem{GP-FL}
G.~Baym and  Ch.~Pethick,
{\it Landau Fermi-liquid Theory}
(Wiley-VCH, Weinheim, 2004, 2nd. Edition).


\bibitem{Mjump}
A.B.~Migdal,
Sov. Phys. JETP 5, 333 (1957); V.M.~Galitsky and A.B.~Migdal,
Sov. Phys. JETP {\bf 7}, 96 (1958); A.B.~Migdal, Sov. Phys. JETP
{\bf 16}, 1366 (1963).



\bibitem{M67}
A.B.~Migdal, {\it Theory of Finite Fermi Systems and Properties
of Atomic Nuclei} (Wiley and Sons, N.Y., 1967);
A.B.~Migdal,
{\it Teoria Konechnyh Fermi System i Svoistva Atomnyh Yader}
(Nauka, Moscow, 1965; 1983) [in Russian].

\bibitem{Pomeranchuk}
I. Ya.~Pomeranchuk, Sov. Phys. JETP {\bf 8}, 361 (1958).

\bibitem{pwave-pairing}
D.~Fay and A.~Layzer,
Phys. Rev. Lett. {\bf 20}, 187 (1968);
M.Yu.~Kagan and A.V~Chubukov,
JETP Lett. {\bf 47}, 614 (1988); M.Yu.~Kagan, {\it Modern Trends
in Superconductivity and Superfluidity} (Springer, Heidelberg,
2013).


\bibitem{Vexp95}
D. N.~Voskresensky,
Phys. Lett. B {\bf 358}, 1 (1995).

\bibitem{Pitaev84}
L. P.~Pitaevskii,
JETP Lett. {\bf 39}, 511 (1984).



\bibitem{V93}
D. N.~Voskresensky,
JETP {\bf 77}, 917 (1993).


\bibitem{Melnikovsky}
L.A.~Melnikovsky,
Phys. Rev. B {\bf 84}, 024525  (2011).

\bibitem{BP12}
G.~Baym and C.J.~Pethick,
Phys. Rev. A {\bf 86}, 023602 (2012).

\bibitem{KV2015}
E.E.~Kolomeitsev and D.N.~Voskresensky, Phys. Rev. C {\bf 91},
025805 (2015).

\bibitem{Kolomeitsev:2015dua}
E.E.~Kolomeitsev and D.N.~Voskresensky,
arXiv: 1501.00731.

\bibitem{RMS} G. R\"opke, L. M\"unchow, and H. Schulz,
Phys. Lett. B {\bf 110}, 21 (1982).

\bibitem{SVB} H.~Schulz, D.N.~Voskresensky, and J.~Bondorf,
Phys.\ Lett.\ B {\bf 133}, 141 (1983).


\bibitem{Margueron:2002wk}
J.~Margueron and P.~Chomaz,
Phys.\ Rev.\ C {\bf 67}, 041602 (2003).

\bibitem{Ravenhall:1983uh}
D.G.~Ravenhall, C.J.~Pethick, and J.R.~Wilson,
  Phys.\ Rev.\ Lett.\  {\bf 50}, 2066 (1983).

\bibitem{Maruyama:2005vb}
T.~Maruyama, T.~Tatsumi, D.N.~Voskresensky, T.~Tanigawa, and S.~Chiba,
Phys.\ Rev.\ C {\bf 72}, 015802 (2005);

\bibitem{SaperFayans}
E.E.~Saperstein, S.A.~Fayans, and V.A.~Khodel, Phys. Part.
Nucl., {\bf 9} 221 (1978);
E.E.~Saperstein and S.V.~Tolokonnikov, JETP Lett. {\bf 68}, 553 (1998).


\bibitem{MSTV90}
A.B.~Migdal, Rev. Mod. Phys. {\bf{50}}, 107 (1978); A.B.~Migdal,
E.E.~Saperstein, M.A.~Troitsky, and D.N.~Voskresensky, Phys.
Rept. {\bf 192}, 179 (1990).


\bibitem{KhodelShaginian}
V.A.~Khodel, V.R.~Shaginyan, and V.V.~Khodel, Phys. Rep. {\bf
249}, 1 (1994).

\bibitem{Khodel:2011dx}
V.A.~Khodel, J.W.~Clark, and M.V.~Zverev,
Phys.\ Atom.\ Nucl.\  {\bf 74}, 1237 (2011).


\bibitem{Voskresensky:2000px}
D.N.~Voskresensky, V.A.~Khodel, M.V.~Zverev, and J.W.~Clark,
Astrophys.\ J.\  {\bf 533}, 127 (2000).


\bibitem{Voskresensky:1982vd}
D.N.~Voskresensky and I.N.~Mishustin,
Sov.\ J.\ Nucl.\ Phys.\  {\bf 35}, 667 (1982).


\bibitem{Sadovnikova}
V.A.~Sadovnikova and M.G.~Ryskin,
Phys. Atom. Nucl. {\bf 64}, 440 (2001); V.A.~Sadovnikova,
Phys. Atom. Nucl. {\bf 70}, 989 (2007);
arXiv:1304.0928 (2013).


\bibitem{AAB} A.I.~Akhiezer, I.A.~Akhiezer, and B.~Barts,
Sov. Phys. JETP {\bf 29}, 1120 (1969).


\bibitem{Pethick-Ravenhall88}
C.J.~Pethick and D.G.~Ravenhall,
Ann. Phys. {\bf 183}, 131 (1988).





\bibitem{Speth:2014tja}
J.~Speth, S.~Krewald, F.~Gr\"ummer, P.-G.~Reinhard, N.~Lyutorovich, and V.~Tselyaev,
Nucl.\ Phys.\ A {\bf 928}, 17 (2014).

\bibitem{Wambach:1992ik}
J.~Wambach, T.L.~Ainsworth, and D.~Pines,
Nucl.\ Phys.\ A {\bf 555}, 128 (1993).


\bibitem{VS:2010gf}
V.V.~Skokov and D.N.~Voskresensky,
Nucl.\ Phys.\ A {\bf 847}, 253 (2010);
D.N.~Voskresensky and V.V.~Skokov,
Phys.\ Atom.\ Nucl.\  {\bf 75}, 770 (2012).


\bibitem{Tatsumi:2002dq}
D.N.~Voskresensky, M.~Yasuhira, and T.~Tatsumi,
Nucl.\ Phys.\ A {\bf 723}, 291 (2003);
T.~Maruyama, T.~Tatsumi, D.N.~Voskresensky, T.~Tanigawa, T.~Endo and S.~Chiba,
Phys.\ Rev.\ C {\bf 73}, 035802 (2006).


\bibitem{Altland-Simons}
A.~Altland and B.~Simons, {\it Condensed Matter Field Theory} (CUP, Cambridge, 2010).

\bibitem{Kopietz} P. Kopietz, e-print arXiv. cond-mat/0605402.

\bibitem{Brovman}
E.G.~Brovman and  Yu.~Kagan,
Sov. Phys. JETP {\bf 36}, 1025 (1972);
E.G.~Brovman and A.~Kholas,
Sov. Phys. JETP {\bf 39}, 924 (1974).


\bibitem{IKV00}
Yu.B.~Ivanov, J.~Knoll, and D.N.~Voskresensky, Nucl. Phys. A
{\bf 672}, 313 (2000).



\bibitem{V84}
D.N.~Voskresensky, Phys. Scripta {\bf 29}, 259 (1984); {\it ibid.} {\bf 47},
333 (1993).


\bibitem{D82}
A.M.~Dyugaev, Sov. JETP {\bf 56}, 567 (1982); Sov. J. Nucl. Phys.
{\bf 38}, 680 (1983).


\end{thebibliography}
\end{document}